# Hearing through Vibrations: Perception of Musical Emotions by Profoundly Deaf People

Anastasia Schmitz[1], Catherine Holloway[1,2], and Youngjun Cho[1,2]

[1]UCL Interaction Centre (UCLIC), University College London, London, United Kingdom
[2]Global Disability Innovation Academic Research Centre (GDI-ARC), Department of Computer Science, University College London, London, United Kingdom

## Abstract
Advances in tactile-audio feedback technology have created new possibilities for deaf people to feel music. However, little is known about deaf individuals' perception of musical emotions through vibrotactile feedback. In this paper, we present the findings from a mixed-methods study with 16 profoundly deaf participants. The study protocol was designed to explore how users of a backpack-style vibrotactile display perceive intended emotions in twenty music excerpts. Quantitative analysis demonstrated that participants correctly identified happy and angry excerpts and rated them as more arousing than sad and peaceful excerpts. More positive emotions were experienced during happy compared to angry excerpts while peaceful and sad excerpts were hard to be differentiated. Based on qualitative data, we highlight the benefits and limitations of using vibrations to convey musical emotions to profoundly deaf users. Finally, we provide guidelines for designing accessible music experiences for the deaf community.

## CCS Concepts
Human-centered computing~Human computer interaction (HCI)~Empirical studies in HCI
Social and professional topics~User characteristics~People with disabilities

## Keywords
Haptics, musical emotion, deafness, hearing loss, tactile-audio feedback

## Research Highlights
• Profoundly deaf people's perception of musical emotions presented through vibrations has been underexplored.
• We conducted a mixed-methods study with 16 profoundly deaf participants to investigate how profoundly deaf individuals experience and perceive musical emotions via vibrotactile feedback.
• We contributed an understanding of the potential and limitations of vibrotactile feedback for conveying intended emotions in music to deaf users.
• We produced design guidelines for future research and developments of more accessible tactile-audio platforms for the profoundly deaf community.

---

Email: anastasia_schmitz@gmx.de, c.holloway@ucl.ac.uk, youngjun.cho@ucl.ac.uk

# 1 Introduction

Music is one of the most powerful triggers of strong emotional experiences and their related cognitive processes [21]. Indeed, it can encourage self-regulation and mood enhancements by conveying and inducing emotions through pitch, timbre, frequency, rhythm and mode (major/minor) [4,46]. It is created via a combination of vocal and/or instrumental sounds, which are typically experienced via auditory perception (i.e. hearing). However, music can often not be enjoyed in this manner by people with hearing impairments. At least 11 million people in the UK are affected by hearing loss [1,2] of which roughly one million individuals have severe to total (i.e. profound) loss of hearing above 81 decibels [1,7]. Even more individuals are expected to experience deafness in the future, due to an ageing population and increasing exposure to loud music/sounds in recreational or work-related contexts [1,2]. Hearing loss increases the risk of developing mental health issues. For example, research in the field has consistently documented that particularly uncorrected hearing loss increases the likelihood of experiencing a poorer quality of life, decreased self-efficacy, social isolation, and feelings of being excluded or depressed [1,5,16]. While listening to music has a strong therapeutic impact on mental health [38], it is often physically inaccessible to individuals who experience hearing loss and particularly to people with congenital and profound deafness [5], who are primarily considered in this paper.

Currently, many profoundly deaf individuals experience sound with the help of sign language interpreters, graphic displays, subtitles, hearing aids, and/or cochlear implants, which can convey certain music features and emotions [25,40]. Despite these advancements, 66% of 41 partially and profoundly deaf individuals reported feeling upset about not being able to fully engage with music in a recent survey [40]. Several possible explanations for this finding exist. For example, the availability of sign language interpreters and even subtitles for music is relatively low and many deaf individuals reject hearing aids or implants, as they perceive them as an invasion of deaf culture [36,49]. Deaf culture supports the view that deafness is part of an individual's identity and should be accepted by society rather than "being cured" [36,49]. Hearing aids and implants also tend to have poor spectral and frequency resolutions, thereby conveying pitch and timbre inaccurately [31]. Furthermore, they occasionally cause uncomfortable experiences with loud sounds, particularly when human voices are involved [25]. Therefore, the most common means of experiencing music frequently fail to provide profoundly deaf individuals with an accurate awareness of musical features and the music's emotional impact as well as meaning [25,31].

Advances in tactile-audio feedback technology, which helps convey a physical dimension of sound, have paved the way for deaf individuals to somatically engage with sound [40]. For example, a commercialised tactile-audio platform SubPac uses subwoofers to convert sound waves into vibrations and allow users to feel music directly on their body [50]. These devices open up new opportunities for radically different types of interaction for the deaf community. In this respect, they can be seen as important in moving forward the new *Disability Interaction* manifesto [26]. They have become a pre-requisite for multiple musical concerts and festivals for the deaf, such as Glastonbury's Deaf Zone or London's DeafRave [30]. However, little is known about whether deaf individuals can distinguish between different types of intended emotions in music via vibrotactile feedback.

Exploring the modality's strengths and limitations for conveying musical emotions is essential to inform the design of assistive vibrotactile technologies. These in turn can improve the user experience and innovate the accessibility of music, particularly for people with profound deafness. In this paper, we explore the interaction of sixteen profoundly deaf individuals with music and the effectiveness of vibrotactile feedback for conveying four emotions (happiness, anger, sadness, peacefulness). This paper contributes:

1. Understanding of how profoundly deaf individuals experience and perceive musical emotions via vibrotactile feedback.
2. Understanding of the potential as well as limitations of vibrotactile feedback for conveying intended emotions in music to deaf users.
3. Design guidelines for future research and developments of more accessible audio-haptic technologies for the deaf population.

## 2 Related Work

Our work is based on research in the fields of emotion, audition, tactile perception and vibrotactile feedback.

### 2.1 Emotion, Audition and Tactile Perception

To understand the potential of vibrotactile feedback for emotion perception in music, we first need to explore the intricate relationship between emotion, audition and touch. According to Russel's circumplex model of affect, emotions are a dynamic physiological response to an eliciting stimulus (e.g. sound) and are based on our stimulus appraisal in regard to valence (positive/negative) and arousal (excitement/boredom) [21,45,55]. The most basic emotions on which psychologists agree are sadness, anger, fear, and happiness [15,18]. These emotions are processed in the limbic system and cerebral cortex of the brain [14]. The thalamus in the brain sends information, received by neurotransmitters, to the amygdala, which can generate emotional reactions before cognitive interpretations of the stimulus take place [28].

As indicated, music and sound can be eliciting stimuli for emotional experiences. Sound waves are defined by their frequency and amplitude. Frequency is the number of sound waves per second, which is responsible for our auditory perception of pitch (high/low). Amplitude is the height of sound waves, which is responsible for our perception of the music's loudness [28]. As sound waves enter the inner ear, hair cells fire and transmit the sound's information to the ear's auditory nerves, which send it to the auditory cortex in the brain via the thalamus [3]. The thalamus can then "contact" the amygdala to elicit an emotional response.

The role between music and emotion perception has been thoroughly documented in the literature. In regard to musical compositions, pace (i.e. tempo) and tonality (i.e. combination of sound frequencies) can impact emotion perception in addition to frequency and amplitude. For instance, Chong et al., [11] demonstrated that loud music is typically associated with high arousal and consequent happiness, anger, or fear whereas quiet music is associated with sadness or peacefulness. High frequencies are commonly related to positive emotions whereas low frequencies and atonal music are generally related to more negative emotions. The researchers showed that participants with music experience and those without it were able to determine intended emotions in music. The results have been confirmed by Vieillard et al. [54], who exposed hearing participants to 56 piano excerpts, each defined by a particular combination of sound frequencies, mode, tempo and pitch. Participants rated happy and angry music as highly arousing. Angry music was perceived as negative whereas happy excerpts had a positive emotional valence. Peaceful and sad music both encouraged low arousal although peaceful music was considered as positive whereas sad music was perceived as negative [54].

Individuals who are profoundly deaf and do not use hearing aids or cochlear implants can have similar emotional reactions to music as the hearing individuals in the above studies. However, their reactions predominantly depend on their haptic perception of bass notes and the music's tempo [43]. For example, when a profoundly deaf person stands close to amplified speakers, the air can act as a conductor of sound waves so that the person may feel the sound vibrations on his/her skin via tactile feedback. This effect is possible, as our skin contains four mechanoreceptive afferent units, which react distinctively to different types of sensory information [32]. Research has also revealed that vibrotactile feedback is processed in deaf individuals' auditory cortices and acts as a substitute for sound [47]. It is therefore likely that deaf individuals experience similar emotions in response to music as hearing people.

### 2.2 Vibrotactile Feedback and Emotions in Music

Several prototypes of vibrotactile displays (e.g. EmotiChair [33], Haptic Chair [40], Haptic Cushion [10], The Model Human Cochlea [34]) and commercially available wearable devices (e.g. SubPac backpack, Sound Shirt, Woojer wrist strap) have been proposed to make music more accessible to the deaf community. A recent study has demonstrated that a wearable vibrotactile display (i.e. SubPac) can enhance users' gaming and music experiences by increasing suspense and realism [17]. Vibrating floors can allow deaf individuals to dance in synchrony with music with no performance differences to hearing people [48,52]. Furthermore, vibrating chairs have been positively received by hearing impaired users [30,40]. Such interfaces have also allowed deaf filmmakers to accurately edit music into films [6]. Similarly, Araujo et al. [4] have shown that deaf participants could identify whether a music excerpt was related to a video based on the vibrations' rhythm and energy from a vibrating chair and bracelet.

A few studies have explicitly investigated whether vibrotactile feedback has the potential to convey intended emotions. Yoo et al. [57] presented 24 hearing participants with varying combinations of sound amplitudes, frequencies, durations, and envelope frequencies (i.e. variation of sound waves over time), using tactile icons on a smartphone. Similar to auditory perception, vibrations, caused by increasing amplitudes, increased participants' arousal and emotional valence was strongly dependent on frequency (i.e. increasing frequency = increasingly positive emotions). However, many vibrations were perceived as negative and increased arousal. Thus, tactile feedback could be impractical for conveying low-arousal emotions (e.g. sadness, peacefulness). Wilson and Brewster [55], who applied vibrations to 12 hearing participants' hands, confirmed that high amplitudes decreased perceived emotional valence. Furthermore, long-lasting and intense vibrations (1000ms) increased participants' arousal. However, particularly high frequency vibrations of 200-300Hz were perceived as negative rather than positive.

Karam and colleagues [34] investigated the effectiveness of a vibrotactile chair, called "The Model Human Cochlea". It converts music into vibrations via eight voice coils, aligned at the chair's back. The researchers artificially deafened hearing participants via noise-cancelling earplugs and headphones. Subjects were interviewed and rated their enjoyment, arousal, and valence in response to happy, angry, fearful and sad music. As expected, they enjoyed happy music significantly more than angry or fearful music. However, happiness and sadness were not distinguishable on the valence-spectrum and arousal for sadness did not differ from other emotions. Thus, although some distinctions between musical emotions were possible, hearing participants seemed unable to truly feel intended emotions when using the chair.

### 2.3 Problem Statement and Research Question

Despite promising findings emerging from the literature, no conclusions can yet be drawn on the effectiveness of vibrotactile feedback for conveying intended emotions in music to a profoundly deaf user and perception by the user population. The most relevant studies [34,55,57] involved hearing participants, limiting their generalizability to deaf individuals who tend to have a higher tactile sensitivity and can be expected to differentiate sounds and musical emotions better when exposed to vibrations [41]. Furthermore, existing studies have not investigated whether participants were able to recognize specific emotions after experiencing vibrations and, thus, whether emotions were not only subconsciously experienced but also consciously perceived [54]. Finally, the focus of the literature has been mainly on creating and designing vibration feedback and prototypes, not on assessing the effectiveness of commercially available solutions.

Given this, this paper investigates the use of a commercially available wearable vibrotactile display to answer our research question: To what extent can profoundly deaf individuals perceive music-induced emotions through vibrotactile feedback? Our three hypotheses are:
H1. Happy and angry music will encourage greater arousal than sad and peaceful music.
H2. Happy and peaceful music will be rated as more positive than sad and angry music.
H3. Participants will recognize the intended emotion more often than unintended emotions in each condition (e.g. recognize anger most frequently in angry music).
In addition, this study explores the number of correctly recognized emotions as well as participants' confidence in their judgment across all music conditions.

## 3 Method

### 3.1 Overview

This section presents our study and the research methods, which were designed to understand how profoundly deaf individuals interact with music and experience its intended emotions through vibrotactile feedback. We used a within-subjects design. The independent variable was the type of intended emotion in each of twenty music excerpts, with four conditions: happy, angry, sad and peaceful. The three primary dependent variables were the participants': a) arousal, b) emotional valence, and c) identified emotions in the music. Two further a priori exploratory dependent variables were: d) participants' confidence in their identified emotion, and e) the number of correctly recognized emotions.

## 3.2 Apparatus

Figure 1 illustrates the experimental setup. A commercial vibrotactile backpack (SubPac M2x) was chosen for this study due to its high popularity [50] and as it has been explored in previous HCI research [17]. The device transduces sound frequencies of 1-200Hz into vibrations via two voice coils, which contain vibrotactile membranes. The vibrations are transmitted to the user's back. A default intensity of 50% was used and the device was connected to a MacBook Air laptop via a 3.5mm stereo aux cable. ITunes was used to play twenty piano excerpts in the range between 11 and 16 seconds each. We used music excerpts by Vieillard et al. [54], which were correctly recognized most frequently by their participants. Four further excerpts from their study were used as demonstrations in a practice session. Happy music was composed in major mode with medium-high pitch and tempo (92-196 Metronome Markings, MM). In contrast, angry music was played in minor mode with dissonant, out-of-key tunes and varying tempo (44-172MM). Sad excerpts were composed in minor mode and were slow in tempo (40-60MM) and peaceful music was in major mode with moderate tempo (54-100MM). A pair of JBL Duet NC headphones was used during a practice task.

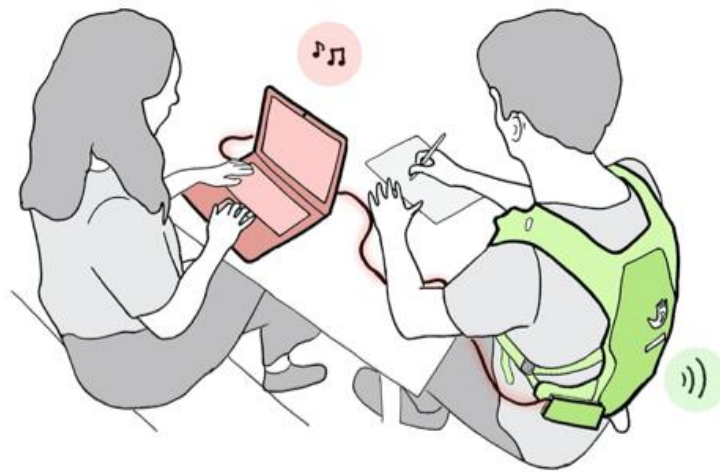

**Figure 1: The experimenter played music on a laptop computer, which was connected to the vibrotactile display worn by the participant.**

## 3.3 Participants

Using a convenience sampling method, 16 profoundly deaf British-born adults (12 male, 4 female), aged 19 to 63 (mean = 39; SD = 14.08), were recruited via social media advertisements to related community-sites. 15 participants wore hearing aids and/or cochlear implants. Individuals with vision impairments, epilepsy, heart problems, anxiety or sensitivity to vibrations were excluded from participation. The exclusion criteria guaranteed that all individuals could independently take part in the experiment. Additionally, they ensured that adverse effects of the presented music and vibrations, such as stress/anxiety and musicogenic seizures (i.e. seizures in response to particular sound frequencies) [19], were avoided. Each participant received an information sheet at the start of the experiment and was asked to sign a consent form before individually taking part. The below study protocol was approved by the Ethics Committee of University College London Interaction Centre and participants were verbally debriefed.

### 3.4 Study Protocol

The 30-minute study was conducted in a meeting room at University College London. It contained a practice task, including vibrotactile feedback as well as audio. This task was followed by a quantitative evaluation phase, which incorporated only vibrotactile feedback and a simultaneous questionnaire exercise. The study then concluded with a qualitative evaluation phase.

As part of the practice trial, the introductory excerpts were played, using the vibrotactile display and a pair of headphones. Immediately after participants switched off their hearing aids or cochlear processors, the introductory excerpts were repeated without headphones. The practice trial was designed particularly for this study to provide participants with context to the device and experiment. It allowed them to differentiate between their auditory and somatic perception of the music's emotion. None of the excerpts' intended emotions were revealed to participants to avoid bias.

During the quantitative evaluation task, the main 20 music excerpts were presented to each participant. Each excerpt was played twice with no break in-between. The researcher, who sat a meter apart from the participant, manually controlled the iTunes playlist. A new excerpt was played after the participant stopped writing and at least 10 seconds after the end of another excerpt to allow individuals to physically, emotionally, and cognitively relax and to distinguish between the music. The excerpts' order was randomized across participants to control for order and learning effects. Participants were asked to simultaneously fill out a questionnaire, adapted from previous research [54,55]. It presented participants with a multiple-choice question per excerpt, which required them to encircle the emotion they recognized in the music from a choice of four options (happy, angry, sad, peaceful). It also included three 9-point Likert-scales per excerpt, assessing participants' arousal (9 = excited, 1 = bored), emotional valence (9 = positive, 1 = negative), and confidence that their identified emotion was the intended one in the music (9 = certain, 1= uncertain). Pictures representing each Likert-point were adapted from a previous study [35].

Following the quantitative evaluation, participants were presented with an open-ended qualitative questionnaire. It inquired about their everyday methods of engaging with music, their previous experience with vibrotactile feedback, and their methods of identifying emotions with the study's vibrotactile display. It also inquired about perceived strengths and weaknesses of vibrotactile technologies and design recommendations. A 9-point Likert-scale further assessed participants' perceived ease of identifying emotions via the vibrotactile feedback (1= very difficult, 9=very easy).

## 4 Results: Quantitative Evaluation

Mean arousal, valence, and confidence values for each music condition were computed per participant. The number of correctly recognized emotions and the frequency with which each emotion was recognized in each music condition were counted for each participant. Arousal and valence ratings were analyzed using a multivariate analysis of variance (MANOVA), as the data met the assumptions: data were within-subjects, continuous, normally distributed, as shown by the Kolmogorov- Smirnoff test, and had more cases in each cell than dependent variables. The data also met assumptions of linearity, multivariate normality, and displayed no multicollinearity as well as no significant multivariate outliers, as indicated by Mahalanobis Distance [51]. Arousal, valence and confidence ratings were also analysed with individual Repeated Measures (RM) ANOVAs. The data met the additional assumptions of normality as well as sphericity and did not display significant univariate outliers. A Friedman test was performed on emotion recognition values given the nonparametric data. The dependent variable was ordinal and one group of participants (within-group design) was measured across more than three conditions [42]. Following significant Friedman tests, post-hoc Wilcoxon signed-rank tests were used with a Bonferroni adjusted value of $p<.01$ to control for multiplicity and consequent Type 1 errors.

### 4.1 Arousal and Valence

The MANOVA demonstrated significant differences between the four music conditions concerning arousal and valence ratings, $F(6,88)=15.74$, $p<.001$, Wilks' $\lambda=.23$, $\eta_{P2}=.52$. Two univariate post-hoc analyses, using RM

ANOVAs, were conducted in regard to this data. There was a significant difference in valence ratings between music conditions, $F(3,13)=6.53$, $p=.006$, Wilks' $\lambda=.40$, $\eta_{p2}=.60$, regarding happy and angry music ($p=.002$). However, no significant differences existed between sad and happy ($p=1.00$), sad and peaceful ($p=1.00$), happy and peaceful ($p=.94$), angry and sad ($p=.68$), as well as angry and peaceful music conditions ($p=.50$).

Significant differences existed between music conditions for arousal ratings, $F(3,13)=24.88$, $p<.001$, Wilks' $\lambda=.15$, $\eta_{p2}=.85$. Pairwise comparisons showed significant differences in participants' arousal between happy and sad ($p=.001$), happy and peaceful ($p<.001$), angry and sad ($p<.001$), as well as angry and peaceful music ($p<.001$). However, there were no significant differences between sad and peaceful ($p=1.00$) as well as happy and angry music conditions ($p=.128$). Means and standard deviations are provided in Table 1 and Figure 2 contains a graphic visualization of the results.

Table 1: Mean values for arousal, valence, and confidence when listening to happy, angry, sad, and peaceful music. Within columns, means followed by the same letter are significantly different from one another.

| Music Condition | Arousal Ratings (M, SD) | Valence Ratings (M, SD) | Confidence Ratings (M, SD) |
|---|---|---|---|
| Happy | 6.40 (1.42)[bc] | 6.06 (1.28)[a] | 6.30 (1.39) |
| Angry | 7.20 (1.00)[de] | 4.88 (1.15)[a] | 6.05 (1.62) |
| Sad | 3.79 (1.35)[bd] | 5.61 (1.19) | 6.16 (1.64) |
| Peaceful | 4.09 (1.29)[ce] | 5.54 (.96) | 5.66 (1.90) |

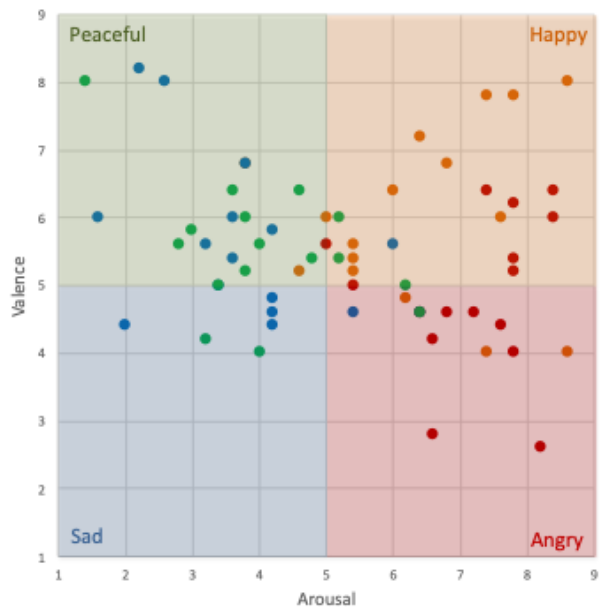

**Figure 2: Each participant's mean arousal and valence ratings for happy (orange), angry (red), sad (blue) and peaceful (green) music. The four quadrants emphasize where ratings of the same color would be expected.**

## 4.2 Confidence

A one-way RM ANOVA on participant's confidence ratings showed that there were no significant differences between the four music conditions in relation to participant's certainty that their identified emotion was the intended one of the music, $F(3,13)=.75$, $p=.28$, Wilks' $\lambda=.75$, $\eta_{p2}=.25$. Respective means and standard deviations are summarized in Table 1.

## 4.3 Identified Emotions Across Music

Four Friedman tests were conducted to explore differences across music conditions in the number of identified happy, angry, sad, and peaceful emotions. See Table 2 for median (Md) values and Figure 3 for a graphic visualization. There was a significant difference in the number of times happiness was identified across all four music conditions, $\chi_2(3, 16)=30.07$, $p<.001$. A post-hoc Wilcoxon signed-rank test demonstrated no significant differences in the small number of times happiness was recognized between sad and peaceful music ($Z=-.584$, $p=.559$, $r=.07$). However, happiness was significantly more often identified in happy compared to angry ($Z=-3.18$, $p=.001$), sad ($Z=-3.44$, $p=.001$, $r=.43$), and peaceful music ($Z=-3.35$, $p=.001$, $r=.42$). It was also more often recognized in angry compared to sad ($Z=-2.81$, $p=.005$, $r=.35$) and peaceful music ($Z=-2.55$, $p=.011$, $r=.32$). There were also significant differences in the number of times that anger was identified across music conditions, $\chi_2(3, 16)=32.71$, $p<.001$. A post-hoc Wilcoxon signed-rank analysis showed that there were no significant differences between sad and happy music ($Z=-2.34$, $p=.02$, $r=.29$), peaceful and happy ($Z=-2.57$, $p=.01$, $r=.32$), and peaceful and sad music ($Z=.00$, $p=1.00$). However, anger was significantly more often identified in angry compared to happy ($Z=-2.91$, $p=.004$, $r=.36$), sad ($Z=3.44$, $p=.001$, $r=.43$) and peaceful music ($Z=-3.45$, $p=.001$, $r=.43$).

Table 2: Number of identified emotions across music conditions. Within columns, means followed by the same letter are significantly different from one another.

| Music condition | Identified as Happy (Md) | Identified as Angry (Md) | Identified as Sad (Md) | Identified as Peaceful (Md) |
|---|---|---|---|---|
| Happy | 3.00[abc] | 1.00[f] | .00[ik] | .00[mo] |
| Angry | 1.00[ade] | 3.00[fgh] | .00[jl] | .00[np] |
| Sad | .00[bd] | .00[g] | 2.00[ij] | 2.00[mn] |
| Peaceful | 0.50[ce] | .00[h] | 1.00[kl] | 2.50[op] |

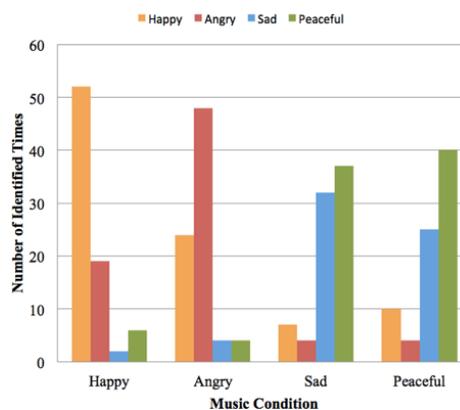

Figure 3: The total number of participants' identified emotions across the four music conditions.

A Friedman test on "sadness" identifications across music conditions also demonstrated significant differences, $\chi_2(3, 16)=24.76$, $p<.001$. A post-hoc Wilcoxon signed-rank test showed that there were no significant differences between angry and happy ($Z=-1.00$, $p=.317$, $r=.13$) and, again, peaceful and sad music ($Z=-1.23$, $p=2.18$, $r=.15$). Sadness was recognized in sad music significantly more often compared to happy ($Z=-3.10$, $p=.002$, $r=.39$) and angry music ($Z=-3.08$, $p=.002$, $r=.39$). However, it was also recognized more often in peaceful compared to happy ($Z=-2.84$, $p=.005$, $r=.36$) and angry music ($Z=-2.71$, $p=.007$, $r=.34$).

Finally, there were significant differences in the frequency with which peacefulness was identified across all music conditions, $\chi_2(3, 16)=28.16$, $p<.001$. A post hoc test showed similar results to the previous analysis on sadness identifications. There were no significant differences between angry and happy ($Z=-.82$, $p=.414$, $r=.10$), and peaceful and sad music ($Z=-.42$, $p=.677$, $r=.05$). However, peacefulness was more frequently identified in sad than happy ($Z=-2.95$, $p=.003$, $r=.37$) and angry music ($Z=-3.16$, $p=.002$, $r=.40$). It was also more often recognized in peaceful compared to happy ($Z=-3.35$, $p=.001$, $r=.42$) and angry music ($Z=-3.32$, $p=.001$, $r=.42$).

### 4.4 Correct Recognitions

The number of correct recognitions across all four music conditions was also examined to explore whether some conditions were more/less difficult for recognizing emotions than others (see Table 2 and Figure 3). Although the Friedman Test pointed out significant differences, $\chi_2(3, 16)=9.04$, $p=.03$, the post hoc Wilcoxon signed-rank tests revealed no differences between music conditions after applying the Bonferroni correction. Thus, participants correctly identified emotions to a similar extent between happy and angry ($Z=-.53$. $p=.60$, $r=.07$), happy and sad ($Z=-2.13$, $p=.033$, $r=.27$), happy and peaceful ($Z=-1.29$, $p=.20$, $r=.16$), sad and angry ($Z=-2.11$, $p=.04$, $r=.26$), sad and peaceful ($Z=-.80$, $p=.42$, $r=.10$), as well as angry and peaceful music ($Z=-.99$, $p=.32$, $r=.12$).

## 5 Results: Qualitative Evaluation

14 participants responded to the qualitative questionnaire. The remaining two participants arrived late to their experimental session and had insufficient time to complete the questionnaire. Four individuals reported previous experience with vibrotactile technology.

The average perceived ease of identifying emotions in the study's music excerpts (see Figure 4) was 5 (i.e. neither easy nor difficult). This finding mimics participants' relatively low confidence in their identified emotions in the quantitative evaluation session, as demonstrated in Table 1. Participants' responses to open-ended questions on the qualitative questionnaire were analyzed via inductive Thematic Analysis [8]. Two completed self-report questionnaires were randomly selected and coded, forming the basis to an initial codebook. The codes were then applied holistically across the remaining 12 questionnaires and were updated iteratively. Another researcher, unfamiliar with the research and study participants, reviewed all codes in 50% of randomly selected transcripts and agreed with 86% of them. Disagreements were resolved over discussion and the final codebook contained 36 codes, which were summarized into five themes: a) typical music experience, b) vibrotactile feedback benefits, c) vibrotactile feedback limitations, d) emotion identification through vibrotactile feedback, and e) design recommendations.

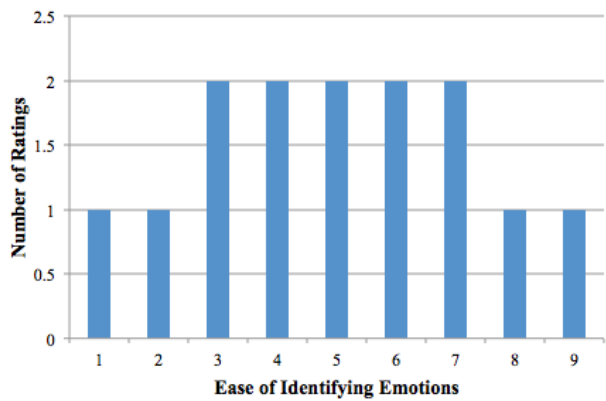

**Figure 4: Participants' ease of identifying emotions in experimental music excerpts, rated on a 9-point Likert scale (1=very difficult, 5= neither easy nor difficult, 9=very easy).**

## 5.1 Typical Music Experience

Participants reported five methods of engaging with music in their everyday life. They experienced music by: a) wearing headphones (N=8), b) turning up their speakers' volume while wearing hearing aids or cochlear processors (N=8), c) feeling vibrations through speakers/instruments (N=9), d) displaying song lyrics (N=1), and/or c) streaming music directly to their hearing aids (N=3), for example, via the Bluetooth neckloop (P1) or via "*smart digital aids, which live stream music through an app on my phone directly to my aids*" (P2). Although methods a-c were most commonly reported, participants outlined several challenges with using them. Participant 3 wrote: "*I have to have the volume turned up almost to maximum to hear music. This is quite anti-social for my family, so I tend to use earphones and then the music is still super quiet*", whilst participant 9 mentioned: "*I also have a large speaker on my desk and feel the bass through that, but that seriously disturbs my neighbours*".

Thus, participants' most accessible means of engaging with music currently come at a cost for their immediate surroundings (i.e. family, partners, friends, neighbors) and frequently promote negative self-awareness. This limitation supports our claim that new forms of assistive technology for the deaf community are needed. Only two participants own a vibrotactile display (SubPac M2 backpack), which they use daily in combination with headphones or hearing aids, as "*SubPac boosts the music, but I still rely on my hearing aids*" (P12).

## 5.2 Vibrotactile Feedback Benefits

Most participants (N=11) reported that the clear vibrotactile feedback from the wearable vibrotactile display resulted in an engaging and enhanced music experience in the study. Three participants further emphasized that the vibrations increased their enjoyment of the music. This finding is in line with previous research, which demonstrated greater enjoyment of music through vibrations [40]. For example, participant 9 reported that "*the vibrations are a lot clearer when sent to the back*". Participant 16 claimed that the vibrotactile device enabled him to "*feel at one with the music*", and participant 5 experienced "*a big enhancement in music appreciation*".

Furthermore, four participants reported that the vibrotactile feedback acts as a protection of their remaining hearing abilities by replacing loud sounds travelling through hearing aids or cochlear processors in particularly noisy environments. For instance, participant 1 wrote "*you don't have to worry about hurting your ears, listening to really loud music*". Participant 2 further imagined "*how it can be useful to convey some music elements in especially loud setting where HA users tend to switch off our devices like concerts or generally crowded and noisy environments*".

Participants also emphasized the relaxing qualities of feeling vibrations (N=3). For example, participant 13 reported that she would use them "*to help switch off at night-time*". Further reported strengths were the comfort and portability of wearable vibrotactile displays (N=2) and their impact on the inclusion of deaf individuals in music-related social activities (N=1).

## 5.3 Vibrotactile Feedback Limitations

Despite the outlined benefits, participants reported several limitations of using vibrotactile feedback to convey musical emotions. Most participants (N=12) suggested that vibrations provided insufficient information to distinguish between emotions in music, as: a) only bass frequencies can be adequately transduced into vibrations, b) differences between music features appear too subtle due to the narrow presentation of sound frequencies, and c) additional features, such as song lyrics, can currently not be conveyed solely through vibrations. Indeed, participant 9 mentioned that the SubPac M2x "*misses out major parts of the music. What about text? What about subtle differences in cross-tones?*" Participant 13 further claimed that "*it would take a while to decipher what it is trying to tell you, as it really focuses on the bass rather than the high frequencies*".

In light of the above limitations, some participants (N=4) emphasized the need for training (i.e. learning how to interpret vibrations). For example, participant 9 proposed the use of "*a manual that helps me understand the meaning of vibrations (where they occur, duration etc.)*".

Many participants (N=7) mentioned that the vibrations and vibrotactile device in the study were physically and aesthetically uncomfortable, predominantly as the display was considered heavy and large. Other complaints included tooth (N=1) and skin sensitivities (N=1) to vibrations. As such, participant 2 "*did not like the intense deep-layered vibrations from the SubPac, as it seemed to travel through my body up to my teeth and created an uncomfortable sensation*". Participant 7 reported that the "*skin feels itchy after the use due to constant vibrations*". The wide range of factors, which caused discomfort, need to be considered for the design of wearable haptic displays, as they emphasize the large variation in user responses to vibrations and the importance of customization.

Finally, the relatively high cost of most commercially available vibrotactile interfaces was mentioned as a limitation (N=2). They were perceived as "*prohibitively expensive for most deaf users, as unemployment is exceptionally high within the community*" (P5).

## 5.4 Emotion Identification through Vibrotactile Feedback

Sad and peaceful music excerpts were predominantly identified based on low-intensity vibrations (N=7). They also contained "*long vibrations without interruption*" (P1) and slow tempo (N=13). However, participants mentioned that peaceful excerpts were difficult to identify and distinguish from sad excerpts (N=6), as reflected by the study's quantitative findings. Nevertheless, eight participants reported that peaceful music was "*slightly faster than sad music and little bit more arousing (heavier beats)*" (P16). Therefore, distinctions between both emotions in music were possible but had to be learnt and required strong attention to detail. Additionally, some participants may have struggled to identify the two emotions, as their judgments were dependent on their mood before music was played (N=6). For instance, participant 13 wrote: "*What should have possibly been sad felt peaceful because of my happy mood*".

Peacefulness (N=3), sadness (N=3), and happiness (N=3) were also identified in music excerpts based on regular vibration intervals whereas an unpredictable rhythm and overlapping vibrations were representative of anger (N=3). As such, participant 13 reported that both sad and peaceful music were "*slow and the beats felt quite clear (separate from one another) …. Happiness felt like clear definite beats… Anger felt like a mismatch of different beats and tones at the same time*". Similarly, participant 3 wrote that "*if the music seemed fraught or sporadic or unpredictable, then it felt more negative*" (P3).

Most participants distinguished between angry and happy music based on the vibrations' intensity and tempo. Indeed, participants (N=9) claimed that angry excerpts' "*beat and tempo was fast and very strong and uncomfortable*" (P16). Fast, short, and light or medium vibrations were indicative of happiness (N=7). For example, participant 8 mentioned, "*happy ones just felt quicker and with light, short vibrations*".

## 5.5 Design Recommendations

All surveyed participants proposed that vibrations should complement sound through live streaming hearing aids or corresponding headphones. This was well expressed by participant 3 – "*I don't think I would choose the vibration-only option, because I found this was an enhancer rather than a replacer of music*". This view of vibrations as an enhancer was further developed as participants (N=8) discussed how they would like vibrations to be extended to other regions of the body, such as the entire body or the wrists. These vibrations would help in enabling a more nuanced music experience, which captures the richness of sound and allows a clear distinction between sound frequencies. For example, participant 5 describes how "*whole-body vibrations might also help in distinguishing the lighter vibrations from each other and feel them more intensely*".

In light of the above recommendation, two users would prefer if wearable vibrotactile displays conveyed particular sounds to different body parts to help them with the distinction of sound frequencies and emotions. Participant 9 mentioned that it "*would be good to have (...) some chords only appear in my right shoulder whereas others appear in my left.*" Participant 14 suggested to "*add movements*" of the vibrations to vibrotactile devices. A further two participants suggested that "*it would be nice if SubPac M2x would have an option to help users to locate the sound*" (P7) and three individuals reported that higher frequencies should be conveyed through light vibrations.

Experience modalities were not restricted to vibrations. One person mentioned the potential of using complementary thermal feedback to convey emotions (N=1) whilst many (N=8) would like to experience vibrations combined with visual feedback. Using colors as metaphorical representations of emotions was the most popular choice. For example, participant 7 described how "*colours are easily associated with emotions (blue for sadness, red for anger, green for happiness and purple for peacefulness)*". Participant 9 proposed that colours should be further enhanced with text: "*Lights! Especially to represent moods and emotions (darker ones for negative emotions and light ones for positive emotions). Also, text is important when using just vibrations for music with lyrics!*". As is evident by the number of proposed forms of feedback to be integrated with vibrations, different people would like to experience sound in different ways. This is in part due to the large range of needs of the deaf community, which is highly diverse in regard to individuals' level of hearing impairment and sensory preferences. Therefore, wearable vibrotactile displays should be customizable (N=8).

## 6 Discussion

Based on recent developments of vibrotactile technology, which promise to make music more inclusive and immersive [24], this study investigated the effectiveness of a commercially available wearable vibrotactile display in translating musical emotions into vibrations. Our study has shown how profoundly deaf individuals perceive musical emotions through vibrations and has outlined the potential and limitations of such technologies for conveying intended emotions to users.

### 6.1 Deaf Users' Musical Emotion Perception via Vibrations

Participants rated happy and angry music as more arousing than peaceful and sad music, supporting our first hypothesis. Although participants experienced more positive emotions during happy compared to angry music, they did not experience emotions of differing valence across the remaining conditions, similarly to Karam et al. [34], leading to a rejection of our second hypothesis. Indeed, participants in this study attributed a positive emotional valence to most music stimuli. This finding contrasts with previous research in which hearing participants perceived most vibrations as slightly negative [57]. Therefore, although profoundly deaf individuals appear to perceive vibrations as positive overall, vibrotactile feedback could not make participants truly feel all intended musical emotions. Participants were able to differentiate happy and angry music and could distinguish both from sad and peaceful excerpts. However, the latter two could not be told apart. Even though this finding does not support our third hypothesis, the large effect sizes imply that sad and peaceful music could be distinguishable in larger samples [42].

There were no significant differences between the four conditions in regard to participants' confidence as well as the number of times that emotions were recognized correctly in music. However, the relatively low significance values and the small to medium effect sizes again suggest that particularly happy and sad as well as sad and angry music could be distinguished better in larger samples.

Our findings had consistently high effect sizes, which suggests high validity and allowed us to make informed interpretations of the results [12]. Additionally, the study had a high internal validity and controlled extraneous variables. For instance, a key strength was the study's inclusion of participants with similar degrees of deafness. It also used a popular and commercially available device as well as validated methods and materials [35,50,54,55]. Due to randomizing the music excerpts' order, we also controlled for order and learning effects.

Overall, our study has demonstrated that vibrotactile technologies have potential for conveying music-related emotions. However, valence may need to be conveyed more clearly to evoke intended emotions, particularly for music that encourages low arousal via slow or moderate tempo (i.e. sad and peaceful music).

### 6.2 Design Guidelines

Profoundly deaf individuals in this study considered the sole use of vibrations insufficient to convey the richness of sound and musical emotions. The importance of customizable vibrotactile displays and the incorporation of further experience modalities was also emphasized. In line with participants' feedback, we therefore suggest that vibrotactile devices should be available with complementary speakers or live streaming options to hearing aids/cochlear implants. Although profoundly deaf individuals cannot fully experience the richness of audio, people with hearing aids or cochlear implants preferred to use it in conjunction with vibrations.

Secondly, we suggest considering multimodality in designing such music *physical listening* devices that are based on vibrotactile feedback. One is visual feedback (e.g. colour representation, light intensity). The complementary feature could better represent emotions and is in line with suggestions from previous research [27,40]. Indeed, previous studies indicated that visuals activate deaf individuals' auditory brain regions and could therefore act as a further substitution for sound [22]. For example, saturated bright colours along the red-yellow spectrum are intuitively perceived as happy whereas cyan-blue colours are related to fear [13]. Also, kinaesthetic feedback (i.e., awareness of movement and force generated by the muscles) can be considered to compensate for tactile feedback (i.e. felt by mechanoreceptors) [23,29].

Lastly, if vibrotactile devices transduced frequencies above 200Hz and were available as different clothing/accessory items (e.g. full-body suits, tops, bracelets), they could provide users with a more nuanced understanding of the music, which has also been suggested by Nanayakkara et al. [40]. It will be important to include deaf individuals in the design and research of future vibrotactile devices to meet their needs [20,37]. Additionally, manuals or demonstrations should explain the meaning of certain vibration patterns and intensities to users.

### 6.3 Limitations and Future Directions

There is room for improvement. The controlled nature of the lab experiment reduced its ecological validity; hence, it would be interesting and insightful to investigate this in different contexts and environmental settings (e.g. musical concerts [30], in-car entertainment [56]). Also, given our focus on profoundly deaf individuals who typically engage with synthesized sounds via hearing aids or cochlear implants, our findings may not be generalizable to the hard-of-hearing population (who use naturally occurring sounds alongside vibrations and born-deaf individuals who have never used hearing aids or cochlear processors and, thus, have different interpretations of music [53]. Indeed, recent research has shown that children with congenital hearing loss were significantly worse at identifying intended emotions in music than children with acquired hearing loss and children with normal hearing [39]. Future research should therefore differentiate between levels of deafness when exploring the effectiveness of vibrotactile feedback to gain a more nuanced understanding of its potential and limitations as well as the needs of different user populations. Lastly, physiological signatures (e.g. respiration [9], heart rate variability [59], vasoconstriction [60]) will be measured in our future studies in order to perform a more in-depth assessment of emotional responses to vibrotactile feedback and provide better insights into vibrations' psychophysiological effects [44,58].

# 7  Conclusion

This study has contributed to research in the field of assistive technologies for the deaf community by conducting a mixed-methods study to explore profoundly deaf individuals' emotion perception in music. It has provided novel insights into deaf users' interaction with music via vibrotactile feedback and the latter's potential as well as limitations for conveying intended emotions. We conclude that vibrotactile feedback is promising to make music more accessible to the profoundly deaf community. However, although participants could distinguish music based on arousal, they were not certain of their judgments and struggled to identify the music's valence. Distinguishing between peaceful and sad music was particularly challenging. Therefore, the effectiveness of additional modalities, such as visual and kinaesthetic feedback, needs to be explored further in future.